\documentclass[11pt]{article}
\usepackage{moriond}
\usepackage[english]{babel}
\usepackage[dvips]{graphicx}

\let\Otemize =\itemize
\let\Onumerate =\enumerate
\let\Oescription =\description
\def\Nospacing{\itemsep=0pt\topsep=0pt\partopsep=0pt\parskip=0pt\parsep=0pt}
\def\Topspac{\vspace{-0.5\baselineskip}}
\def\Botspac{\vspace{-0.2\baselineskip}}
\newenvironment{Itemize}{\Topspac\Otemize\Nospacing}{\endlist\Botspac}
\newenvironment{Enumerate}{\Topspac\Onumerate\Nospacing}{\endlist\Botspac}

\newcommand{\lsim}{\,{\buildrel < \over {_\sim}}\,}
\newcommand{\gsim}{\,{\buildrel > \over {_\sim}}\,}
\newcommand{\sqrtsNN}{\sqrt{s_{\scriptscriptstyle{{\rm NN}}}}}

\newcommand{\gev}{\mathrm{GeV}}
\newcommand{\tev}{\mathrm{TeV}}
\newcommand{\fm}{\mathrm{fm}}

\newcommand{\mum}{\mathrm{\mu m}}

\newcommand{\PbPb}{\mbox{Pb--Pb}}

\newcommand{\pt}{p_{\rm t}}
\renewcommand{\d}{{\rm d}}
\newcommand{\dEdx}{{\rm d}E/{\rm d}x}
\newcommand{\dNdy}{{\rm d}N_{\rm ch}/{\rm d}y}

\newcommand{\Jpsi} {\mbox{J\kern-0.05em /\kern-0.05em$\psi$}\xspace}

\begin{document}
\vspace*{4cm}
\title{HEAVY FLAVOURS IN HEAVY-ION COLLISIONS AT THE LHC:\\ALICE PERFORMANCE}

\author{ A. DAINESE }

\address{Universit\`a degli Studi di Padova and INFN, 
        via Marzolo 8, 35131 Padova, Italy\\
        e-mail: andrea.dainese@pd.infn.it\\
        {\rm on behalf of the ALICE Collaboration~\footnote{
        Presented at the XXXIX$^{\rm th}$ Rencontres de Moriond on QCD and
        High-Energy Hadronic Interactions.}}}
\maketitle

\abstracts{
We present the latest results on the ALICE performance for heavy-quark 
production and quenching measurements, focusing in 
particular on charm particles.
}

\section{Introduction: heavy quarks as probes of QCD matter}
\label{intro}

The ALICE experiment~\cite{tpalice} will study nucleus--nucleus (AA)
collisions at the LHC, with a centre-of-mass energy $\sqrtsNN=5.5~\tev$ per
nucleon--nucleon (NN) pair for the \PbPb~system, 
in order to investigate the properties of QCD matter at energy densities of 
up to several hundred times the density of atomic nuclei. In these conditions
a deconfined state of quarks and gluons is expected to be formed.
As we shall detail in the following paragraphs, heavy quarks are sensitive 
probes of such a medium.  

Heavy-quark pairs ($\rm Q\overline Q$) 
are produced in the early stage of the collision in 
primary partonic scatterings with large virtuality $Q$ and, thus, 
on temporal and 
spatial scales, $\Delta t\sim \Delta r\sim 1/Q$, which are
sufficiently small for the production to be 
unaffected by the properties of the medium.
In fact, the minimum virtuality $Q_{\rm min}=2\,m_{\rm Q}$ 
in the production 
of a $\rm Q\overline Q$ pair implies a space-time scale 
of $\sim 1/(2\,m_{\rm Q})\simeq 1/2.4~\gev^{-1}\simeq 0.1~\fm$ 
(for charm), to be compared
to the expected life-time of the deconfined state at the LHC, 
$\gsim 10~\fm$. 
Thus, the initially-produced heavy quarks experience the full 
collision history. 

Hard partons are regarded as probes of the medium as they are 
expected to lose energy by gluon radiation while propagating 
through high-density QCD matter.\cite{gyulassywang,bdmps,wiedemann}
The attenuation (quenching) of leading hadrons observed in Au--Au collisions 
at RHIC~\cite{david} 
is thought to be due to such a mechanism.
Due to the large values of their masses, the charm and beauty quarks are
qualitatively different probes with respect to light partons, 
since the `dead-cone effect' is expected to reduce the in-medium energy loss
of massive partons.\cite{dokshitzerkharzeev,nestor}
Therefore, a comparative study of the attenuation of massless and 
massive probes at the LHC will allow to test the consistency 
of the interpretation 
of quenching effects as due to energy loss in a deconfined medium and to
further investigate the properties (density) of such a medium.

\section{Heavy-flavour detection in ALICE}
\label{alice}

\begin{figure}[!t]
  \begin{center}
  \includegraphics[width=0.85\textwidth]{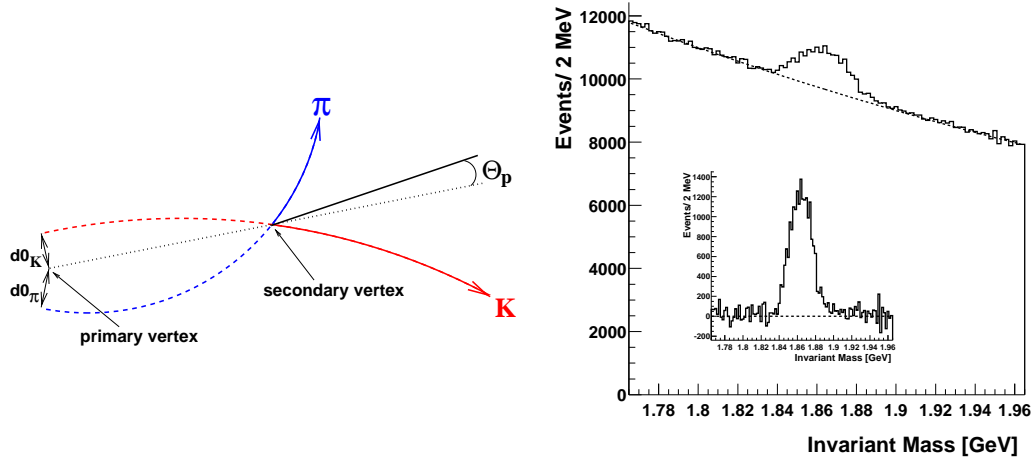}
  \caption{Schematic representation of the $D^0 \to K^-\pi^+$ decay (left). 
           $K\pi$ 
           invariant-mass distribution corresponding to $10^7$ central Pb--Pb 
           events (right); the background-subtracted distribution is shown 
           in the inset.}
\label{fig:D0combined}
\end{center}
\end{figure}

The ALICE experimental setup was designed in order to allow the detection
of $D$ and $B$ mesons in the high-multiplicity environment 
of central \PbPb~collisions at LHC energy, where up to several thousand 
charged particles might be produced per unit of rapidity. 
The heavy-flavour capability of the ALICE detector is provided by:
\begin{Itemize}
\item Tracking system; the Inner Tracking System (ITS), 
the Time Projection Chamber (TPC) and the Transition Radiation Detector (TRD),
embedded in a magnetic field of $0.5$~T, allow track reconstruction in 
the pseudorapidity range $|\eta|<0.9$ with a momentum resolution better than
2\% for $\pt<10~\gev/c$ 
and an impact parameter~\footnote{The impact parameter,
$d_0$, is defined as the distance of closest approach of the track to the 
interaction vertex, in the plane transverse to the beam direction.} 
resolution better than 
$60~\mum$ for $\pt>1~\gev/c$, mainly provided by the two layers of pixel 
detectors of the ITS.
\item Particle identification system; charged hadrons are separated via 
$\dEdx$ in the TPC and in the ITS and via time-of-flight measurement in the 
Time Of Flight (TOF) detector; electrons are separated from pions in the 
Transition Radiation Detector (TRD); muons are identified in the forward muon 
arm covering the range $2.5<\eta<4$. 
\end{Itemize}

Detailed simulation analyses,\cite{D0jpg,D0epjc} 
based on a realistic description of the experimental effects and of the 
background sources, have shown that ALICE has a good potential to carry out 
the comparative quenching studies mentioned in Section~\ref{intro}.
In Section~\ref{charm} we describe the expected performance for the exclusive
reconstruction of $D^0\to K^-\pi^+$ 
decays and the estimated sensitivity for the study 
of charm energy loss. In Section~\ref{beauty} we present the first results
on the possibility to detect $B$ mesons in the inclusive $B\to e^\pm+X$ 
channels. In both studies a multiplicity of $\dNdy=6000$
was assumed for central \PbPb~collisions.
We report the results corresponding to the 
expected number of events collected by 
ALICE per LHC year: $10^7$ for central \PbPb~and $10^9$ for pp collisions. 
The strategies for the measurement of $D$ and $B$ mesons in the
muon arm are currently being optimized; we will not discuss these topics here.

\section{Measurement of charm production and in-medium quenching}
\label{charm}

\begin{figure}[!t]
  \begin{center}
  \includegraphics[width=0.49\textwidth]{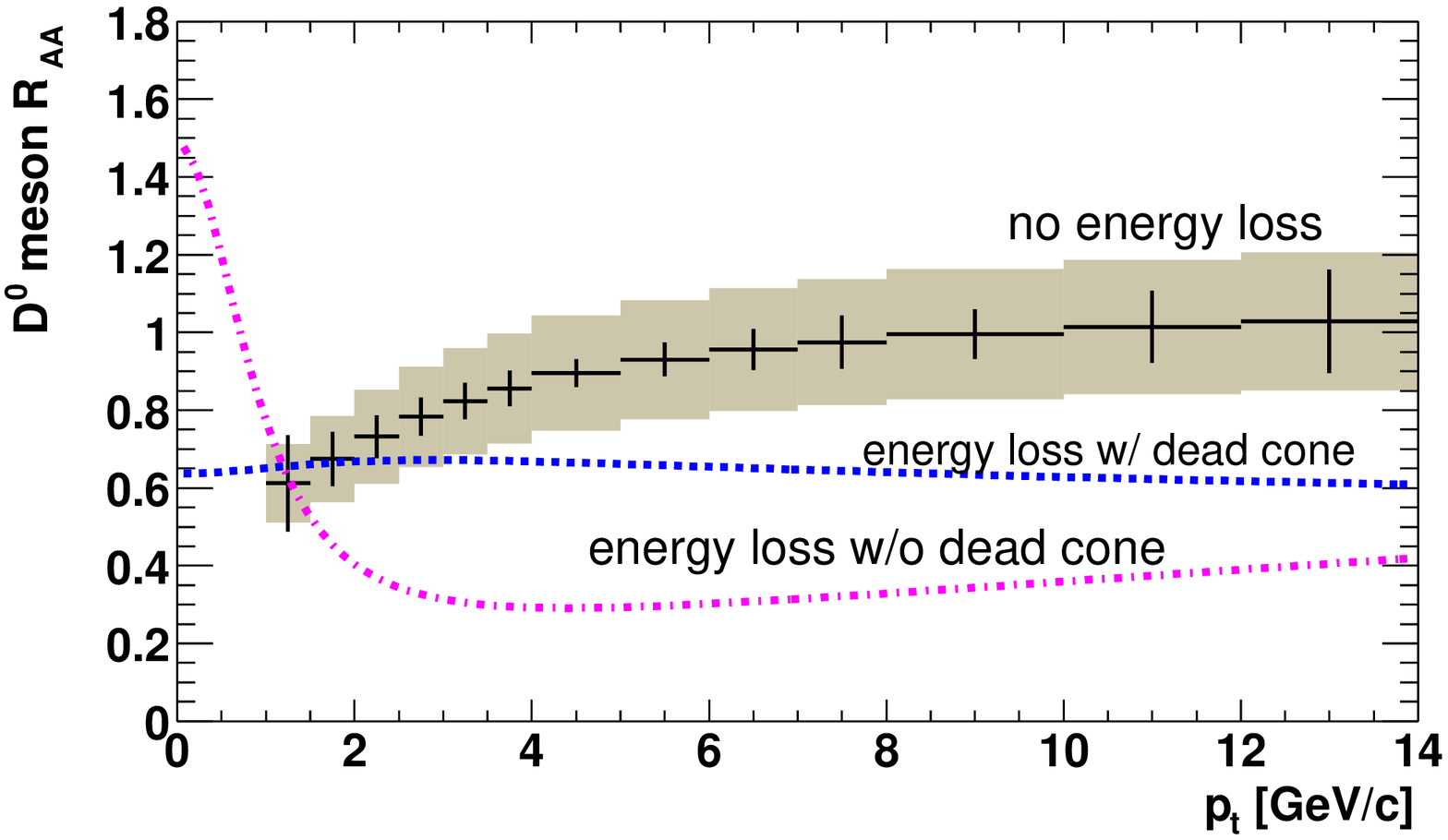}
  \includegraphics[width=0.49\textwidth]{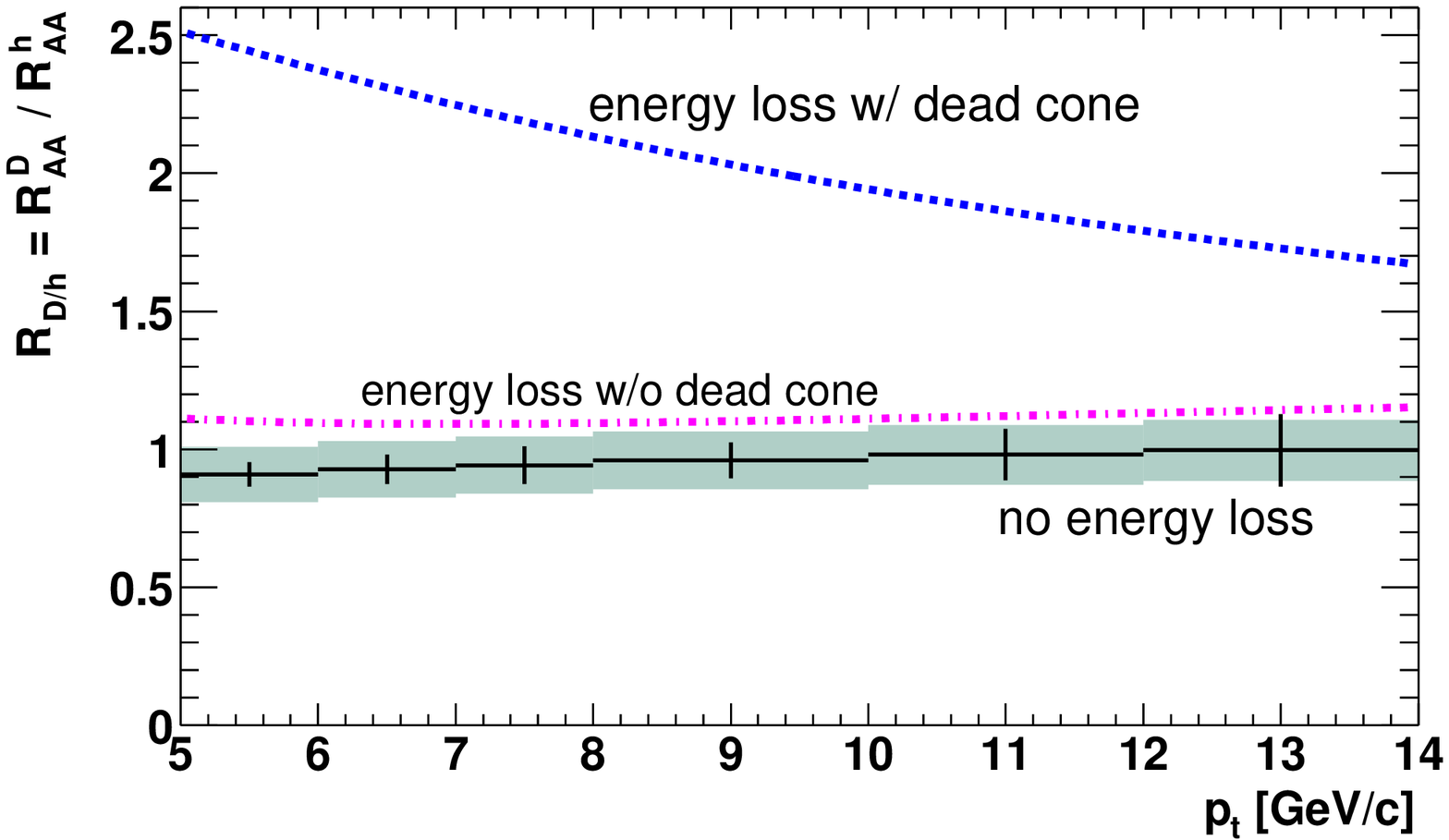}
  \caption{Nuclear modification factor for $D^0$ mesons (left) and 
           ratio of the nuclear modification factors
           for $D^0$ mesons and for charged hadrons (right). Both are shown 
           without and with dead-cone effect. Errors corresponding to 
           the case `no energy loss' are reported: bars = statistical, 
           shaded area = systematic.}
\label{fig:D0quench}
\end{center}
\end{figure}

One of the most promising channels for open charm detection is the 
$D^0 \to K^-\pi^+$ decay (and its charge conjugate) which 
has a branching ratio of about $3.8\%$.
The expected production yields ($\d N/\d y$ at $y=0$), 
estimated~\cite{noteHVQ} on the basis 
of next-to-leading order pQCD calculations, 
for $D^0$ (and $\overline{D}^0$) mesons decaying in a $K^\mp\pi^\pm$ pair 
in central 
Pb--Pb ($5\%~\sigma^{\rm tot}$) at $\sqrtsNN=5.5~{\rm TeV}$ and in pp 
collisions at $\sqrt{s}=14~{\rm TeV}$ are $5.3\times 10^{-1}$ and 
$7.5\times 10^{-4}$ per event, respectively.

Figure~\ref{fig:D0combined} 
(left) shows a sketch of the decay: the main feature 
of this topology is the presence of two tracks with impact parameters 
$d_0\sim 100~\mum$. The detection strategy~\cite{D0jpg} to cope with
the large combinatorial background from the underlying event is based on:
\begin{Enumerate}
\item selection of displaced-vertex topologies, i.e. two tracks with 
large impact parameters
and small pointing angle $\Theta_{\rm p}$ 
between the $D^0$ momentum and flight-line
(see sketch in Fig.~\ref{fig:D0combined});
\item identification of the $K$ track in the TOF detector;
\item invariant-mass analysis (see $\pt$-integrated
distribution in \PbPb~after selections in Fig.~\ref{fig:D0combined}).
\end{Enumerate}
This strategy was optimized separately for pp and \PbPb~collisions, as a 
function of the $D^0$ transverse momentum.\cite{thesis} 
The accessible $\pt$ range is $1$--$14~\gev/c$ for \PbPb~and 
$0.5$--$14~\gev/c$ for pp, with a statistical error better than 15--20\% 
and a systematic error 
(acceptance and efficiency corrections, 
centrality selection for \PbPb) better than 20\%. More details 
are given in Ref.~\cite{thesis}.

We studied~\cite{D0epjc} the sensitivity for a comparison of the energy loss 
of charm quarks and of massless partons by considering:
\begin{Itemize} 
\item the {\it nuclear modification factor} of 
$D$ mesons as a function of $\pt$
\begin{equation}
\label{eq:raa}
  R_{\rm AA}^D(\pt)\equiv
    \frac{{\rm d}N^{\scriptscriptstyle D}_{\rm AA}/{\rm d}\pt/{\rm binary~NN~collisions}}
       {{\rm d}N^{\scriptscriptstyle D}_{\rm pp}/{\rm d}\pt},
\end{equation}
which would be equal to 
1 if the AA collision was a mere superposition of independent 
NN collisions, without nuclear or medium effects; 
\item the ratio of the nuclear modification factors of $D$ mesons and
of charged hadrons:
\begin{equation}
\label{eq:RDh}
   R_{D/h}(\pt)\equiv R_{\rm AA}^D(\pt)\Big/R_{\rm AA}^h(\pt).
\end{equation}
\end{Itemize}

Medium-induced parton energy loss was simulated using
the `quenching weights',\cite{qw} an approximation of the dead-cone 
effect for charm quarks and a Glauber-model based description of the 
collision geometry to calculate in-medium parton path lengths. The density 
of the medium was estimated in order to have $R_{\rm AA}^h\approx 0.2$--$0.3$,
similarly to that measured at RHIC.\cite{david}

The results for $R_{\rm AA}^D$ and $R_{D/h}$ 
are presented in Fig.~\ref{fig:D0quench}. The reported uncertainties are 
discussed in detail in Refs.\cite{D0epjc,thesis}. The effect of nuclear
shadowing, introduced via the EKS98 parameterization,\cite{EKS98} 
is clearly visible 
in the $R_{\rm AA}$ without energy loss for $\pt\lsim 7~\gev/c$. Above this
region, only parton energy loss is expected to affect the nuclear modification 
factor of $D$ mesons. The relative importance of the energy-loss and 
dead-cone effects can be disentangled using the $R_{D/h}$ ratio, which can be
measured with good sensitivity as it is a double ratio 
\mbox{(AA/pp)\,/\,(AA/pp)} (many systematic uncertainties cancel out).
We find that this ratio is enhanced, with respect to 1, only by the dead cone
and, consequently, it appears as a very clean tool to investigate and quantify
this prediction of QCD. 

\section{Perspectives for the detection of beauty in the semi-electronic 
decay channels}
\label{beauty}

The production of open beauty can be studied by detecting the 
semi-electronic decays of $B$ mesons. 
Such decays have a branching ratio of $\simeq 10\%$.
The expected yield ($\d N/\d y$ at $y=0$) for $B\to e^{\pm}+X$ 
in central ($5\%~\sigma^{\rm tot}$) 
Pb--Pb collisions at $\sqrtsNN=5.5~\tev$ is $9\times 10^{-2}$ per event.

The main sources of background electrons are: (a) decays of $D$ mesons; 
(b) decays of light mesons (e.g. $\rho$ and $\omega$);
(c) conversions of photons in the beam pipe or in the inner layers of the
ITS and (d) pions identified as electrons. 
Given that electrons from beauty have impact parameter $d_0\simeq 500~\mum$
and a hard momentum spectrum, we expect to obtain a high-purity sample with a 
strategy that relies on:
\begin{Enumerate}
\item electron identification with a combined $\dEdx$ (TPC) and transition
radiation (TRD) selection, which allows to reduce the pion contamination 
by a factor $10^4$;
\item impact parameter cut to reject misidentified pions and electrons
from sources (b) and (c);
\item transverse momentum cut to reject electrons from charm decays. 
\end{Enumerate}
As an example, with $d_0>180~\mum$ and $\pt>2~\gev/c$, the expected statistics
of electrons from $B$ decays is $5\times 10^4$ for $10^7$ central Pb--Pb
events, with a contamination of about 10\%, mainly given by the decays  
$D\to e^\pm+X$ and $B\to D+X\to e^\pm+X+X'$.
The sensitivity on the extraction of the $b\overline{b}$ production 
cross section and of the $B$-meson $\pt$ distribution is currently being 
investigated.

\section*{Acknowledgements}

The author acknowledges fruitful discussions with F.~Antinori, N.~Armesto, 
A.~Morsch, G.~Paic, K.~$\check{\mathrm{S}}$afa$\check{\mathrm{r}}$\'\i k, 
C.A.~Salgado, R.~Turrisi and U.A.~Wiedemann.

\end{document}